\documentclass[a4paper,prl,twocolumn,showpacs,nofootinbib,superscriptaddress,floatfix]{revtex4-1}
\usepackage{graphics,graphicx}
\usepackage{amsmath, amssymb}
\usepackage{multirow}
\usepackage{color}
\usepackage{verbatim}
\usepackage{hyperref}
\usepackage[normalem]{ulem}  
\usepackage{ulem}
\usepackage{color}
\usepackage{cancel}
\usepackage{mathtools}
\usepackage{epstopdf}

\renewcommand\sout{\bgroup\color{blue} \ULdepth=-.5ex \ULset}
\usepackage[english]{babel}

\begin{document}

\title{Reaching percolation and conformal limits in neutron stars}

\date{\today}
\author{Micha\l{} Marczenko}
\email{michal.marczenko@uwr.edu.pl}
\address{Incubator of Scientific Excellence - Centre for Simulations of Superdense Fluids, University of Wroc\l{}aw, plac Maksa Borna 9, PL-50204 Wroc\l{}aw, Poland}
\author{Larry McLerran}
\address{
Institute for Nuclear Theory, University of Washington, Box 351550, Seattle, Washington 98195, USA
}
\author{Krzysztof Redlich}
\address{Institute of Theoretical Physics, University of Wroc\l{}aw, plac Maksa Borna 9, PL-50204 Wroc\l{}aw, Poland}
\author{Chihiro Sasaki}
\address{Institute of Theoretical Physics, University of Wroc\l{}aw, plac Maksa Borna 9, PL-50204 Wroc\l{}aw, Poland}
\address{International Institute for Sustainability with Knotted Chiral Meta Matter (SKCM$^2$), Hiroshima University, Higashi-Hiroshima, Hiroshima 739-8511, Japan}

\begin{abstract}
Generating an ensemble of equations of state that fulfill multi-messenger constraints, we statistically determine the properties of dense matter found inside neutron stars (NSs). We calculate the speed of sound and trace anomaly and demonstrate that they are driven towards their conformal values at the center of maximally massive NSs. The local peak of the speed of sound is shown to be located at values of the energy and particle densities which are consistent with deconfinement and percolation conditions in QCD matter. We also analyze fluctuations of the net-baryon number density in the context of possible remnants of critical behavior. We find that the global maxima of the variance of these fluctuations emerge at densities beyond those found in the interiors of NSs. 
\end{abstract}

\maketitle

\section{Introduction}

Neutron stars (NSs) serve as unique extraterrestrial laboratories for probing dense matter. Densities within their core can reach up to several times the nuclear saturation density ($n_{\rm sat}=0.16~\rm fm^{-3}$). It is expected that NSs may contain particles other than nucleons such as hadronic resonances or quarks. A widely held expectation is that if quark degrees of freedom are present then their appearance occurs with a strong first-order phase transition, leads to non-monotonic behavior of the equation of state (EoS), in particular, the speed of sound $c_s^2 = \partial p / \partial \epsilon$, where $p$ is the pressure and $\epsilon$ is the energy density. On the other hand, the transition to new degrees of freedom might not involve a first-order phase transition.

The structure of the sound speed provides valuable insights into the microscopic description of dense matter at densities much larger than the saturation density~(see, e.g.,~\cite{Tan:2021nat, Zhang:2019udy, Tews:2018kmu, Legred:2021hdx}). The speed of sound at zero temperature is expressed as 
\begin{equation}
c_s^2 = \frac{n_B}{\mu_B\chi_B} \textrm,
\end{equation}
where
\begin{equation}
\chi_B = \frac{\partial^2 p}{\partial \mu_B^2} = \frac{\partial n_B}{\partial \mu_B}
\end{equation}
is the second-order cumulant of the net-baryon number density. At small densities ($n_B \lesssim 2~n_{\rm sat}$), EoS is provided reliably by  chiral effective field theory~\cite{Tews:2018kmu} where $c_s^2$ is found to likely violate the conformal value of $1/3$. Asymptotically, $c_s^2$ approaches the conformal limit as dictated by Quantum Chromodynamics (QCD), which is confirmed by perturbative QCD (pQCD) calculations~\cite{Fraga:2013qra}. However, the density range found in NSs is not accessible by these perturbative methods and the structure of the EoS remains unknown. Great progress in constraining the EoS was achieved by systematic analyses of recent astrophysical observations of the massive pulsar PSRJ0740+6620~\cite{Cromartie:2019kug, Fonseca:2021wxt, Miller:2021qha, Riley:2021pdl} and PSR J0030+0451~\cite{Miller:2019cac} by the NICER collaboration, and the constraint from the GW170817 event~\cite{LIGOScientific:2018cki}, within parametric models of the EoS~(see, e.g.,~\cite{Alford:2013aca, Alford:2017qgh, Li:2021sxb, Annala:2019puf, Somasundaram:2021clp, Annala:2017llu}).

The trace anomaly scaled by the energy density
\begin{equation}
\Delta = \frac{1}{3} - \frac{p}{\epsilon} \textrm,
\end{equation}
was recently proposed as a measure of conformality~\cite{Fujimoto:2022ohj}. Due to thermodynamic stability and causality $-2/3 \leq \Delta \leq 1/3$. The speed of sound can be alternatively decomposed in terms of $\Delta$ as
\begin{equation}
c_s^2 = \frac{1}{3} - \Delta - \epsilon \frac{d \Delta}{d \epsilon}\textrm.    
\end{equation}
As the scale invariance becomes restored in QCD, $\Delta\rightarrow0$ and $c_s^2\rightarrow 1/3$. The vanishing of the trace of the stress-energy tensor is a consequence of conformal invariance, and $\Delta$ provides an alternative measure of this property. In~\cite{Fujimoto:2022ohj}, it was argued that $\Delta$ might monotonically approach zero with increasing energy density, and the matter becomes conformal around $1~\rm GeV/fm^3$. As a consequence of the quick approach to conformality, $c_s^2$ develops a peak at lower densities. Such description of dense matter is naturally obtained in quarkyonic models~\cite{McLerran:2007qj, Duarte:2021tsx, Kojo:2021ugu, Kojo:2021hqh, Fukushima:2015bda, McLerran:2018hbz, Jeong:2019lhv, Sen:2020peq, Cao:2020byn, Kovensky:2020xif}. This is in contrast to a description of the transition quark matter involving a first-order phase transition or rapid crossover, where the sound velocity would be expected to have a minimum.

Fluctuations of conserved charges are known to be promising observables for the search of the critical behavior at the QCD phase boundary~\cite{Stephanov:1999zu, Asakawa:2000wh, Hatta:2003wn} and chemical freeze-out of produced hadrons in heavy-ion collisions~\cite{Bazavov:2012vg, Borsanyi:2014ewa, Karsch:2010ck, Braun-Munzinger:2014lba, Vovchenko:2020tsr, Friman:2011pf}. In particular, fluctuations have been proposed to probe the QCD critical point (CP) in the beam energy scan programs at the Relativistic Heavy Ion Collider at Brookhaven National Laboratory and the Super Proton Synchrotron at CERN, as well as the remnants of the criticality at vanishing and finite baryon densities~\cite{Braun-Munzinger:2020jbk, Friman:2011pf, Karsch:2019mbv, Braun-Munzinger:2016yjz}. In general, it is expected that cumulants of conserved charges are sensitive to critical fluctuations. In QCD, chiral CP belongs to the $Z(2)$ universality class; thus, it is expected for the fluctuations to diverge at CP, which makes them useful probes of remnants of critical behavior in the relativistic heavy-ion collisions. It has also been recently argued that at low temperatures and large baryon densities, experimentally measured cumulants may allow for direct measurement of the sound velocity~\cite{Sorensen:2021zme}.

In this work, we demonstrate that matter inside the cores of maximally massive NSs becomes almost conformal. We analyze the properties of the speed-of-sound and the net-baryon number susceptibility in the context of medium composition and possible critical behavior or its remnants. In particular, we link the observed behavior of the speed of sound with the percolation of nucleons and the emergence of quark or quarkyonic matter.

\section{Methods}

We construct an ensemble of EoSs based on the piecewise-linear speed-of-sound parametrization introduced in~\cite{Annala:2019puf}. The model has been already used in several other works~\cite{Annala:2021gom, Altiparmak:2022bke, Ecker:2022xxj}. Here, we follow the prescription provided in Ref.~\cite{Altiparmak:2022bke}. At densities $n_B < 0.5~n_{\rm sat}$, we use the Baym-Pethick-Sutherland (BPS) EoS~\cite{Baym:1971pw}. In the range ($0.5 - 1.1)~n_{\rm sat}$, we use the monotrope EoS, $P = Kn_B^\Gamma$, where $\Gamma\in (1.77,3.23)$ is sampled randomly and $K$ is matched with the BPS EoS at $0.5~n_{\rm sat}$. At densities $n_B \gtrsim 40~n_{\rm sat}$, we use the pQCD results for the pressure, density, and speed of sound of cold quark matter in $\beta$-equilibrium~\cite{Fraga:2013qra}. The density and speed of sound can be calculated straightforwardly from the pressure. In this work, we use the pQCD results down to $\mu_{\rm pQCD}=2.6~\rm GeV$.

At densities $1.1~n_{\rm sat} \leq n_B \leq n(\mu_{\rm pQCD})$ we use the piecewise-linear parametrization of the speed of sound:
\begin{equation}
    c_s^2(\mu) = \frac{\left(\mu_{i+1} - \mu\right)c_{s,i}^2 + \left(\mu - \mu_i\right)c_{s,i+1}^2}{\mu_{i+1} - \mu_i} \textrm,
\end{equation}
where $\mu_i \leq \mu \leq \mu_{i+1}$. We generate $N$ pairs of $\mu_i$ and $c^2_{s,i}$, where $\mu_i \in [\mu(n_0),\mu_{\rm pQCD}]$ and $c_{s,i}^2 \in [0,1]$. The values of $\mu_1$ and $c^2_{s,1}$ are fixed by the values of the monotrope EoS at $n_0=1.1~n_{\rm sat}$, and $\mu_N = \mu_{\rm pQCD}$.

The net-baryon number density can be expressed as
\begin{equation}\label{eq:nb}
    n_B(\mu) = n_0 \exp{\int\limits_{\mu_0}^{\mu}\mathrm{d}\nu \;\frac{1}{\nu\;c_s^2(\nu)}} \textrm,
\end{equation}
where $n_0=1.1~n_{\rm sat}$ and $\mu_0 = \mu(n_0)$. Integrating Eq.~\eqref{eq:nb} gives the pressure:
\begin{equation}
    p(\mu) = p_0 + \int\limits_{\mu_0}^\mu \mathrm{d}\nu\; n_B(\nu) \textrm,
\end{equation}
where $p_0 = p(\mu(n_0))$. In addition to requiring consistency with the pQCD results at high densities, we impose the observational astrophysical constraints. First, we require that the EoSs support the lower bound of the maximum-mass constraint, $M_{\rm TOV}\geq(2.08\pm0.07)~M_\odot$, from the measurement of J0740+6620~\cite{Fonseca:2021wxt}. Second, we utilize the GW170817 event measured by the LIGO/Virgo Collaboration (LVC). Viable EoSs should also result in a $1.4~M_\odot$ NS with tidal deformability of a $1.4~M_\odot$ NS, $\Lambda_{1.4} =190^{+390}_{-120}$~\cite{LIGOScientific:2018cki}. In total, we analyzed a sample of $4.62\times10^5$ EoSs with $N=7$ segments that fulfill the imposed observational and pQCD constraints.

\begin{figure*}
    \centering
    \includegraphics[width=.48\linewidth]{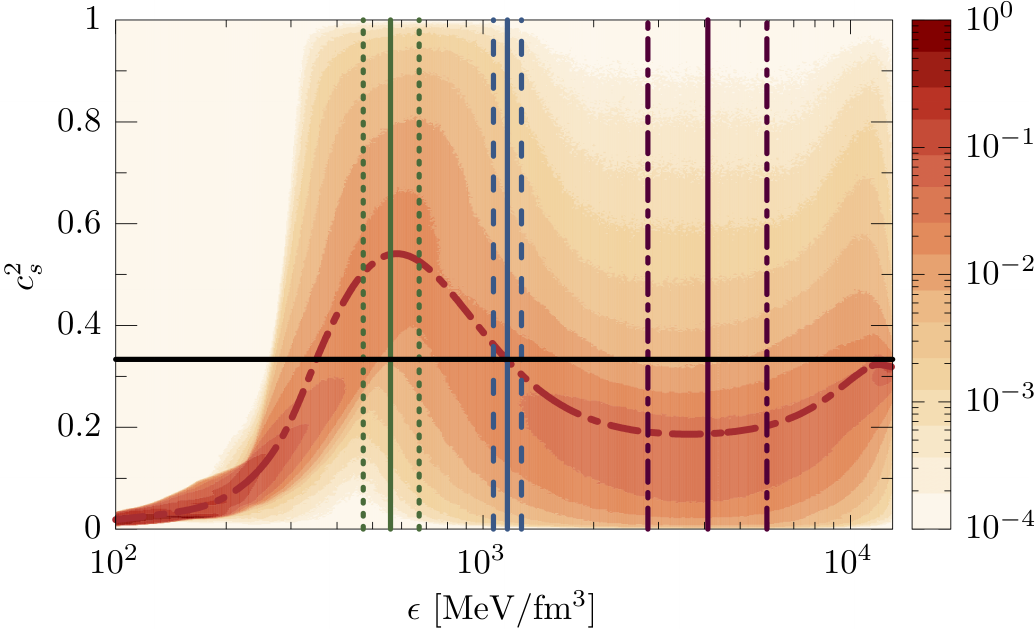}\;\;\;
    \includegraphics[width=.48\linewidth]{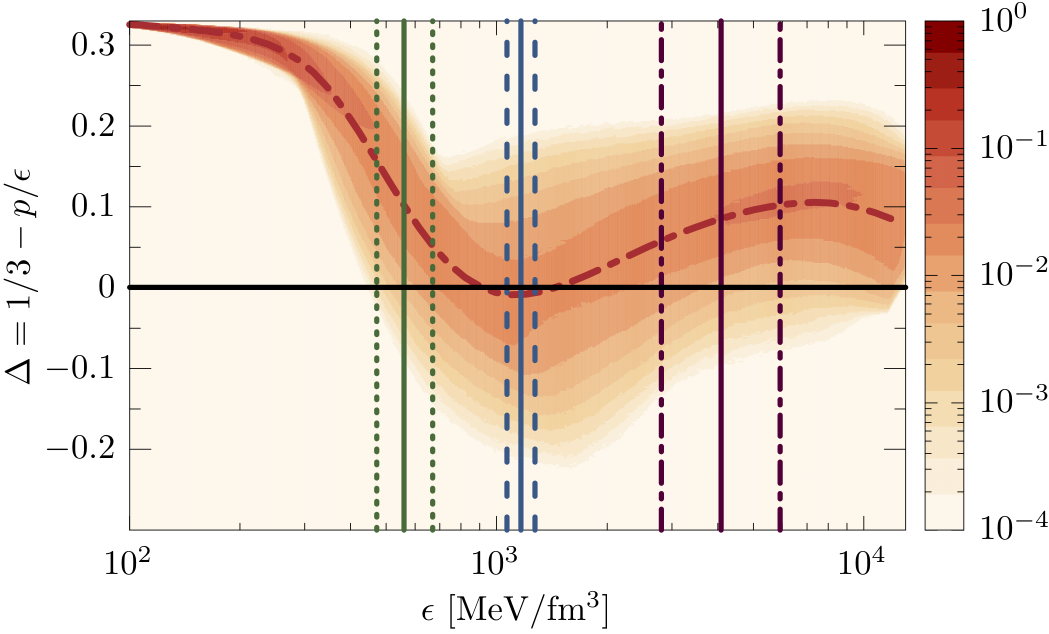}
    \caption{Probability density functions (PDFs) of the speed of sound (left panel) and trace anomaly (right panel) as functions of the energy density. The red, dash-dotted lines show the averages of these quantities. Vertical lines show the median and $1\sigma$ credibility region for the position of the peak in $c_s^2$ (green solid and dotted lines), values at the center of maximally massive NSs (blue solid and dashed lines), and the position of the peak in $\hat\chi_B$ (purple solid and dash-dotted lines). Horizontal, black lines mark the conformal values of $c_s^2=1/3$ (left panel) and $\Delta = 0$ (right panel).}
    \label{fig:cs2_trace_hist}
\end{figure*}

\section{Results}
In the left panel of Fig.~\ref{fig:cs2_trace_hist}, we show the probability distribution function (PDF) of the speed of sound as a function of energy density. The speed of sound swiftly increases at low densities and generates a peak above $1/3$. Notably, the distribution seems to reflect the generic peak-dip structure, and, at large densities, $c_s^2$ converges to the pQCD result. Similar PDF was also obtained in~\cite{Altiparmak:2022bke}. We have performed an average of the speed of sound at all energy densities and integrated it to obtain the pressure. Such EoS yields maximal mass $M_{\rm TOV} = 2.20~M_\odot$ and tidal deformability $\Lambda_{1.4} = 408$, which are in good agreement with astrophysical constraints.

We note, however, that the average may not well reflect individual EoSs. To illustrate this, we have located the local peaks in $c_s^2$ that appear below $\epsilon_{\rm TOV}$ in each EoS in our sample. Their median at $1\sigma$ confidence level is $c_{s, \rm peak}^2 = 0.82^{+0.07}_{-0.08}$ at $\epsilon_{\rm peak} = 0.559^{+0.110}_{-0.088}~\rm GeV/fm^3$. The value of $c_s^2$ differs from the calculated average, where the most plausible value of the peak is $c_s^2 = 0.54$ at $\epsilon = 0.582~\rm GeV/fm^3$. This is because this region is comprised of EoSs that peak below the median and decrease around it, as well as EoSs for which the sound speed keeps increasing in the vicinity of the estimated value for energy density. Nevertheless, the location of the energy density at the peak position of sound velocity stays nearly unchanged. In our sample, we also obtain  $n_{B,\rm peak} = 0.54^{+0.09}_{-0.07}~\rm fm^{-3}$ which corresponds to $\mu_{B,\rm peak}=1.283^{+0.090}_{-0.070}~\rm GeV$.

The behavior of the speed of sound shown in Fig.~\ref{fig:cs2_trace_hist} is very different from that obtained in QCD matter at finite temperature~\cite{Redlich:1985uw, HotQCD:2014kol}. Based on the first principle lattice QCD (LQCD) calculations, it is known that $c_s^2$ never exceeds the conformal limit and exhibits a minimum at the critical energy density $\epsilon_c=0.42\pm 0.06~\rm GeV/fm^{3}$, where chiral symmetry is partially restored and quarks are deconfined~\cite{HotQCD:2018pds, Mykhaylova:2020pfk}. Decrease of $c_s^2$ with energy density towards its minimum from the hadronic side is linked to attractive interactions with resonance formation~\cite{Mykhaylova:2020pfk, Castorina:2009de}. Very different behavior in a cold nuclear matter of $c_s^2$ is due to the dominance of repulsive interactions which implies increasing $c_s^2$ with energy density towards its maximum~\cite{Zeldovich:1961sbr}. Considering that quark deconfinement could be linked to a non-monotonic behavior of $c_s^2$, one can identify the maximum of $c_s^2$ as being due to a phase change from nuclear to quark or quarkyonic matter.

Phenomenologically, deconfinement can be linked to the percolation of hadrons of a given size~\cite{Magas:2003wi, Castorina:2008vu, Satz:1998kg, Fukushima:2020cmk, Braun-Munzinger:2014lba}. Relating the peak in the speed of sound to the percolation threshold in QCD, one can estimate the critical density at which nucleons start to overlap. In percolation theory of objects with constant volume $V_0 = (4/3)\pi R_0^3$, this critical density is given by $n^{\rm per}_c = 1.22/V_0$~\cite{Braun-Munzinger:2014lba}. Recently, the proton mass radius was extracted from the experimental data of $\phi$ photoproduction measured by CLAS~\cite{Dey:2014tfa} and LEPS~\cite{LEPS:2005hax} collaborations. The average from these experiments yields $R_0 = 0.80 \pm 0.05$~fm for $\sqrt{s} \in [2.02, 2.29]$~GeV.~\cite{Wang:2022uch}~\footnote{We note that the proton radius is still not well established and can be as small as $0.55~\rm fm$ (see, e.g., Fig.~9 in Ref.~\cite{Wang:2022uch})}. Consequently, this yields $n^{\rm per}_c = 0.57^{+0.12}_{-0.09}~\rm fm^{-3}$, which is remarkably consistent with $n_{B,\rm peak} = 0.54^{+0.09}_{-0.07}~\rm fm^{-3}$ where $c_s^2$ reaches its maximum.

The extracted parameters of the energy density and particle density corresponding to the percolation threshold in the NS EoS can be compared with the values obtained in hot QCD matter at the chiral crossover temperature $T_{\rm pc}=156.5\pm 1.5~\rm MeV$, where quarks are deconfined. From the discussion above, it is clear that the energy density at the peak position of the speed of sound is of the same order as the LQCD critical energy density $\epsilon_c=0.42\pm 0.06~\rm GeV/fm^{3}$ at deconfinement~\cite{HotQCD:2018pds}. It is interesting to note that $\epsilon_c$ corresponds to the energy density inside the nucleon, $\epsilon_0\simeq m_N/V_0$. Indeed, considering the nucleon mass radius $r_0\simeq 0.8~\rm fm$, one gets $\epsilon_0\simeq 0.44~\rm GeV/fm^{3}$. 

Particle density $n_c$ in QCD matter at $T_{\rm pc}$ can be estimated based on the thermal model analyses of particle production in heavy ion collisions and experimental data~\cite{Andronic:2017pug, Andronic:2018qqt}. There it was shown that in Pb-Pb collisions at $\sqrt s=2.76~\rm TeV$ hadrons are produced at the QCD phase boundary at $T_{\rm pc}$ from the fireball of volume $V=4175\pm 380~\rm fm^{3}$~\cite{Andronic:2017pug, Andronic:2018qqt, Braun-Munzinger:2014lba}. Taking the ratio of number of hadrons per unit of rapidity $N_t=2486\pm 146$, measured by ALICE collaboration, and the above fireball volume, one gets $n_c=0.596\pm 0.065~\rm fm^{-3}$. This value is consistent with the critical percolation density and the extracted density $n_{B,\rm peak}$ at the peak position of the speed of sound.

Following the above discussion, one can conclude that the appearance of the maximum in speed of sound in the interior of NSs can be attributed to the change of medium composition, from hadronic to quark or quarkyonic matter. Thus, purely hadronic NS EoS has limited applicability up to the extracted critical percolation conditions. These are of the same order as found in QCD at finite temperature and vanishing or small baryon densities, where percolation of hadrons corresponds to deconfinement of quarks and partial restoration of chiral symmetry.

We also find the estimate of the median for the largest central energy density found in neutron stars, $\epsilon_{\rm TOV} = 1.163^{+0.107}_{-0.97}~\rm GeV/fm^3$ at $1\sigma$ confidence level. The corresponding value of the speed of sound is $c_{s, \rm TOV}^2 = 0.28\pm0.06$ at $1\sigma$ confidence level, which is remarkably close to the conformal value. We conclude that it is plausible for heavy NSs to feature a peak in the speed of sound in their cores with $c_s^2$ close to unity and which then decreases to $\sim 1/3$. Such non-monotonicity has been recently conjectured~\cite{Kojo:2020krb} and observed in statistical analysis of the EoS~\cite{Ecker:2022xxj}. For instance, the quarkyonic description of dense matter leads to a rapid increase, accompanied by a peak in the speed of sound~\cite{McLerran:2018hbz, McLerran:2007qj, Hidaka:2008yy}.

\begin{figure}[t!]
    \centering
    \includegraphics[width=\linewidth]{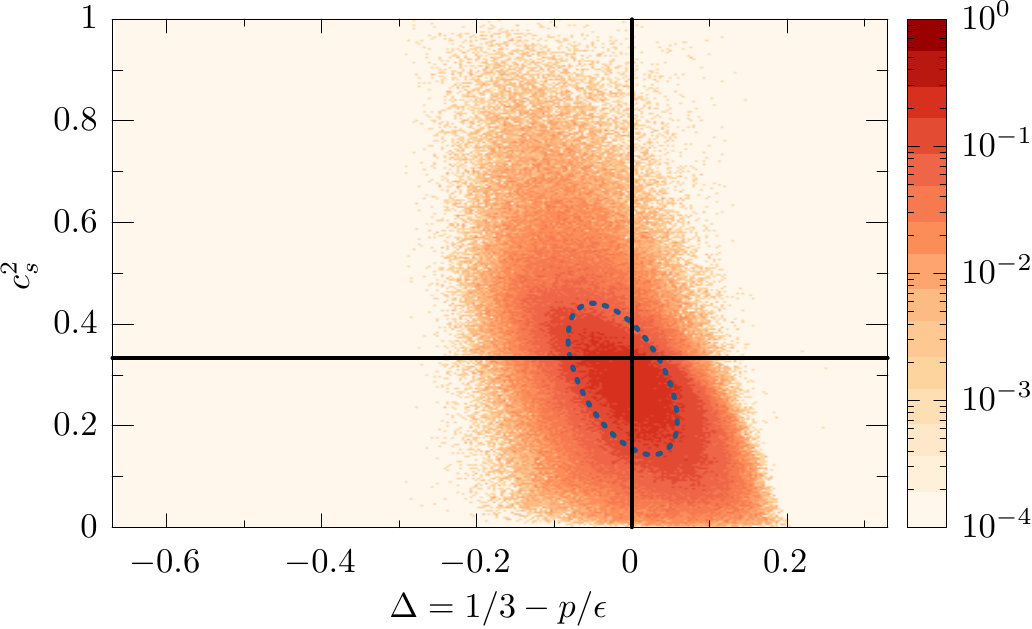}
    \caption{PDF of the trace anomaly vs the speed of sound at the center of maximally stable NSs. The dotted, blue ellipse marks the $1\sigma$ credibility around the mean. The black horizontal and vertical lines mark the conformal values.}
    \label{fig:trace_cs2_maxM}
\end{figure}

In the right panel of Fig.~\ref{fig:cs2_trace_hist}, we show the calculated PDF of the trace anomaly. It decreases monotonically up to $\epsilon \sim\epsilon_{\rm TOV}$, where its average reaches zero,  as proposed in Ref.~\cite{Podkowka:2018gib}. Furthermore, the possibility of the negative trace anomaly at sufficiently high density, as first shown by Zel'dovich~\cite{Zeldovich:1961sbr}, is statistically not excluded. At larger densities, the distribution becomes broader due to the lack of astrophysical constraints and increases to reach the pQCD constraint at positive values. We note that trace anomaly may be monotonic even if the speed of sound develops a peak. This is traced to the derivative part of $\Delta$~\cite{Fujimoto:2022ohj}.

In Fig.~\ref{fig:trace_cs2_maxM}, we plot the PDF of the central values of the speed of sound and trace anomaly obtained for the maximally massive stellar configurations. We report values of the median $c_{s, \rm TOV}^2 = 0.28\pm0.06$ and $\Delta_{\rm TOV} = -0.01\pm0.03$ at the $1\sigma$ confidence level. This is remarkably close to the conformal value and suggests the existence of strongly-coupled conformal matter at the cores of massive NSs. We note that the averaged EoS gives $c_{s,\rm TOV}^2 = 0.42$, $\Delta_{\rm TOV} = -0.01$, which are consistent with our results.

At present one cannot discriminate if $\Delta>0$ at densities above $\epsilon_{\rm TOV}$. Such conjecture would have an immense impact on the allowed mass-radius configurations~\cite{Fujimoto:2022ohj}. Nevertheless, current data seems to allow for the possibility that $\Delta$ might eventually become negative (see Fig.~\ref{fig:cs2_trace_hist}). Also, the sound velocity must approach $1/3$ from below, based on pQCD computations, so that at some high density the sound velocity has a minimum. This minimum corresponds to a peak in the baryon number susceptibility and might be associated with the remnant of a phase transition.

The peak in $c_s^2$ is generated by stiffening of the EoS, but the pQCD EoS requires softening of the EoS at intermediate densities. In general, the onset of new degrees of freedom should be reflected in the EoS by the emergence of critical behavior or its remnants. While a rapid increase in the speed of sound signals significant stiffening of the EoS, it is not necessarily informative of the emergence of some critical behavior. In order to quantify possible critical signals in the EoS, we consider the dimensionless second-order cumulant of the net-baryon density, $\hat\chi_B \equiv \chi_B /\mu_B^2 = n_B / (\mu_B^3 c_s^2)$, and locate its global maximum, $\hat \chi_B^{\rm max}$. Note that we neglect signals from the liquid-gas phase transition at low densities.

It turns out that more than 99\% of the global maxima in $\hat\chi_B$ lie at energy densities above $\epsilon_{1.4}$ and roughly 93\% in the high-density domain beyond the gravitational stability. We find the median $\epsilon_{\rm \chi} = 4.084^{+1.834}_{-1.275}~\rm GeV/fm^3$ at $1\sigma$ confidence level. The onset of criticality which is linked to the maximal baryon-density fluctuations in the EoS is therefore unlikely to be found in the cores of NSs.

\section{Conclusions}

We have statistically determined the NS EoS in view of current observational constraints and analyzed the properties of the speed of sound and trace anomaly. While the sound speed is found to develop a peak above the conformal value of $1/3$, the trace anomaly monotonically decreases towards zero. We have shown that the centers of maximally massive NSs contain matter that is almost conformal. We have argued that the peak in $c_s^2$ can be phenomenologically interpreted by connecting it to the phase boundary obtained in first-principle lattice QCD calculations and percolation threshold extracted from heavy-ion collision experiments. We found that they are remarkably consistent with each other, which strongly indicates a phase change in dense medium. It is noted that the deconfinement of quarks appears under general percolation conditions in both hot and dense matter. We also analyzed fluctuations of the net-baryon number density $\chi_B$ in the context of possible remnants of critical behavior. Considering the location of the global maxima of $\chi_B$ for each EoS, we have found that they are lying in the high-density region beyond the gravitational stability. Thus, when relating the appearance of a global maximum of $\chi_B$ with the remnant of criticality it is very unlikely that it appears in the cores of NSs.

Given the anticipated near-future increase in the accuracy of astrophysical measurements as well as results from forthcoming large-scale nuclear experiments FAIR at GSI and NICA in Dubna, it will become necessary to understand the above statistically identified EoS in terms of fundamental properties of strong interactions described by QCD. Especially, it is vital to formulate the emergence of conformal matter along with the spontaneous breaking of chiral symmetry and its partial restoration inside NSs, as well as the onset of quark or quarkyonic matter. This can be possibly achieved within the application of advanced parity doublet models with the elusive interplay between chiral symmetry breaking~\cite{Marczenko:2021uaj, Marczenko:2022hyt}, deconfinement~\cite{Benic:2015pia, Marczenko:2017huu}, and the topology at the Fermi surface~\cite{Kojo:2021ugu}. We will elaborate on these issues elsewhere.

\paragraph{Acknowledgments.}
This work is supported partly by the Polish National Science Centre (NCN) under OPUS Grant No. 2018/31/B/ST2/01663 (K.R. and C.S.), Preludium Grant No. 2017/27/N/ST2/01973 (M.M.), and the program Excellence Initiative–Research University of the University of Wroc\l{}aw of the Ministry of Education and Science (M.M).
The work of C.S. was supported in part by the World Premier International Research Center Initiative (WPI) through MEXT, Japan.
M.M. acknowledges helpful comments from Tobias Fischer. K.R. also acknowledges the support of the Polish Ministry of Science and Higher Education and stimulating discussions with David Blaschke. The work of L. M. was supported by the U.S. DOE under Grant No. DE-FG02-00ER41132. L. M. gratefully acknowledges the hospitality of the faculty at the University of Wroc\l{}aw for its hospitality during the time this research was initiated.

\bibliographystyle{apsrev4-1}
\bibliography{biblio}

\end{document}